\documentclass[prl,twocolumn]{revtex4}

\usepackage{graphicx}
\usepackage{epstopdf}%
\usepackage{amsmath,amssymb,dsfont,upgreek}

\usepackage{natbib}   

\usepackage[colorlinks=true,  linkcolor=blue, citecolor=blue, urlcolor=blue]{hyperref}

\begin{document}

\title{Test of scaling theory in two dimensions in the presence of valley splitting and intervalley scattering in Si-MOSFETs.}

\author{Alexander Punnoose}
\affiliation{Physics Department, City College of the City University of New York, New York, NY 10031, USA}
\email{punnoose@sci.ccny.cuny.edu} 

\author{Alexander M. Finkel'stein}
\affiliation{Department of Condensed Matter Physics, Weizmann Institute of Science, Rehovot 76100, Israel}
\affiliation{Department of Physics, Texas A\&M University, College Station, TX 77843, USA}

\author{A. Mokashi and S.\ V.\ Kravchenko}
\affiliation{Physics Department, Northeastern University, Boston, Massachusetts 02115, USA}

\begin{abstract}
We show that once the effects of valley splitting and intervalley scattering are incorporated, renormalization group theory consistently describes the metallic phase in silicon metal-oxide-semiconductor field-effect transistors down to the lowest accessible temperatures. \end{abstract}


\maketitle


The two-parameter scaling theory of quantum diffusion in two dimensions\,\cite{punnoose02,punnoose05} has been  remarkably successful in describing the properties around the metal-insulator transition (MIT) in  electron systems confined to silicon inversion layers (MOSFETs)\,\cite{punnoose07,pudalov_RG_JETP,pudalov_RG_PRL}. The theory is based on the scaling hypothesis that both the resistivity and the electron-electron scattering amplitudes become scale dependent in a diffusive system due to the singular long ranged nature of the diffusive propagators, $\mathcal{D}(q,\omega)=1/(Dq^2+\omega)$, in a disordered medium\,\cite{yellowbook,pedagogical}.  
The predicted scale dependencies calculated using renormalization group (RG) theory\,\cite{punnoose02}  were recently verified experimentally in Ref.\,\cite{punnoose07} without any fitting parameters. Since the theory considered the valleys to be degenerate and distinct, the experiments were limited to temperatures larger than the characteristic valley splitting and intervalley scattering rate ($T\agt 500$\,mK). The effects of scaling are, however,  significant at low temperatures and it is therefore important to test the scaling hypothesis at much lower temperatures. 
We show that when the RG theory is extended to include valley splitting and intervalley scattering\,\cite{punnoose09a}  the scaling properties in the metallic phase can be described quantitatively down to the lowest reliably accessible  temperatures,  $T\approx 200$\,mK.

The evolution  with scale (temperature) of the two-parameters, namely, the resistance, $\rho$, and the electron-electron interaction strength, $\gamma_2$,  in the spin-triplet channel  were discussed in detail   for $\rho\alt 1$ (in units of $\pi h/e^2$)  in terms of RG theory in Ref.\,\cite{punnoose02}. (In Fermi-liquid notation, $\gamma_2$ is related to the amplitude $F_0^a$ as $\gamma_2=-F_0^a/(1+F_0^a)$.)   The theory predicts that, %
while $\gamma_2$ increases monotonically as the temperature is reduced, 
$\rho$ behaves non-monotonically, changing from insulating behavior ($d\rho/dT<0$) at high temperatures to metallic behavior ($d\rho/dT>0$) at low temperatures, with the crossover occurring when $\gamma_2$ attains the value $\gamma_2^*=0.45$. Although the maximum value $\rho_{\text{max}}$ occurs at a crossover temperature $T=T_{\text{max}}$, both of which are sample specific and hence non-universal, the two-parameter scaling theory predicts that the behaviors of $\rho(T)/\rho_{\text{max}}$ and $\gamma_{2}(T)$ are universal when plotted as functions of $\xi=\rho_{\text{max}}\ln(T_{\text{max}}/T)$. The above predictions, including the value of $\gamma_2^*$, were verified experimentally in Refs.\,\cite{punnoose02,punnoose07} in the temperature range where the two valleys may be considered to be degenerate and distinct.

For  $n$-(001) silicon inversion layer the conduction band has two almost degenerate valleys located close to the $X$-points in the Brillouin zone. While the sharpness of the interface of the inversion layer leads to the splitting, $\Delta_v$, of the two valley bands, the atomic scale irregularities  found at the interface gives rise to a finite intervalley scattering rate, $\hbar/\tau_\perp$\,\cite{ando}. The singularity of the diffusion modes, especially those in the valley-triplet sector, are  cut-off  at low frequencies as a result\,\cite{fukuyama1,punnoose09a}. 
Hence, the specific form of the RG equations, which is sensitive only to the number of singular modes, depends on  if $k_BT$ is greater than or less than the scales $\Delta_v$ or/and $\hbar/\tau_\perp$. 

The relevant RG  equations for the different temperature ranges may be combined  as follows\,\cite{punnoose09a}:
\begin{subequations}
\begin{eqnarray}
\frac{d\rho}{d\xi}&=&\rho^2\left[1-(4K-1)\left(\frac{\gamma_2+1}{\gamma_2}\log(1+\gamma_2)-1\right)\right]\hspace{0.5cm}\\
\frac{d\gamma_2}{d\xi}&=&\rho\frac{(1+\gamma_2)^2}{2}
\end{eqnarray}\label{eqn:RG}
\end{subequations}
The parameter $K$ accounts for the number of singular diffusion modes in each temperature range.  For temperatures $T\agt T_v$ and $T_\perp$, where $k_BT_v=\Delta_v$ and $k_BT_\perp=\hbar/\tau_\perp$, the two bands are effectively degenerate and distinct; the constant $K$ in this case is proportional to the square of the number of valleys, $n_v$, i.e., $K=n_v^2=4$  ($n_v=2$ for silicon).  In the temperature range $T_\perp\alt T \alt T_v$, the two bands remain distinct but are split and hence each valley contributes independently to $\Delta\sigma(b)$, i.e., $K=n_v=2$. At still lower temperatures $T\alt T_\perp$, intervalley scattering mixes the  two valleys to effectively produce a single valley so that $K=1$. 

A few important clarifications regarding the use of Eq.\,(\ref{eqn:RG}) are discussed below.
First,  for the case $K=2$, when the bands are split but distinct, it has been shown that using a single amplitude $\gamma_2$ to describe the interaction in all the seven $(4K-1)$ modes is an approximation that is valid only if the temperature range $T_\perp \alt T\alt T_v$ is not too wide\,\cite{punnoose09a}.  In general, when the bands are split  certain amplitudes evolve differently from $\gamma_2$, thereby necessitating the need to go beyond the two-parameter scaling description\,\cite{burmistrov_spinvalley,punnoose09a}. The deviation is large  when the RG evolution is allowed to proceed to exponentially large scales or $T\ll T_v$. In our case, however, since $T_\perp$, which effectively mixes the two bands, is only a fraction smaller than $T_v$, the deviation of the amplitudes is quickly limited by $T_\perp$. We therefore assume that all the amplitudes remain degenerate and contribute equally to $\rho$, which amounts to taking $K=2$  in  Eq.\,(\ref{eqn:RG}). 

The second point concerns the weak-localization (WL) contribution\,\cite{aabook} to Eq.\,(\ref{eqn:RG}). It is seen experimentally that the phase breaking rate, $\hbar/\tau_{\varphi}$, saturates at low electron densities ($n\alt 10^{11}$\,cm$^{-2}$) for $T\alt 500$\,K. Correspondingly, a strong suppression of the WL correction is also observed in this regime\,\cite{dephasing_kravchenko}. These observations are consistent with our results, as is discussed later. We have therefore neglected the weak-localization contribution in Eq.\,(\ref{eqn:RG}) when analyzing the cases $K=2$ and $1$ (these are the relevant cases at low temperatures).

In Ref.\,\cite{punnoose07} it was shown that $\gamma_2$ may be determined experimentally  by exploiting the $b^2$ dependence of the magnetoconductance $\Delta\sigma(b)\equiv\Delta\sigma(B,T)=\sigma(B,T)-\sigma(0,T)$ in a weak parallel magnetic field  $b= g\mu_B B/k_B T\alt 1$. In the weak field limit $\Delta\sigma(b)$ is given as\,\cite{tvr,castellani98}
\begin{equation}
\Delta \sigma(b)= - 0.091\frac{e^2}{\pi
h}K\gamma_2\left(\gamma_2+1\right)b^2\label{eqn:linear}
\end{equation}
Hence the slope of $\Delta\sigma(b^2)$ provides a direct measure of $\gamma_2$, given of course that $K$ is known.

\begin{figure}[htb]
    \centering
        \includegraphics[width=.45\textwidth]{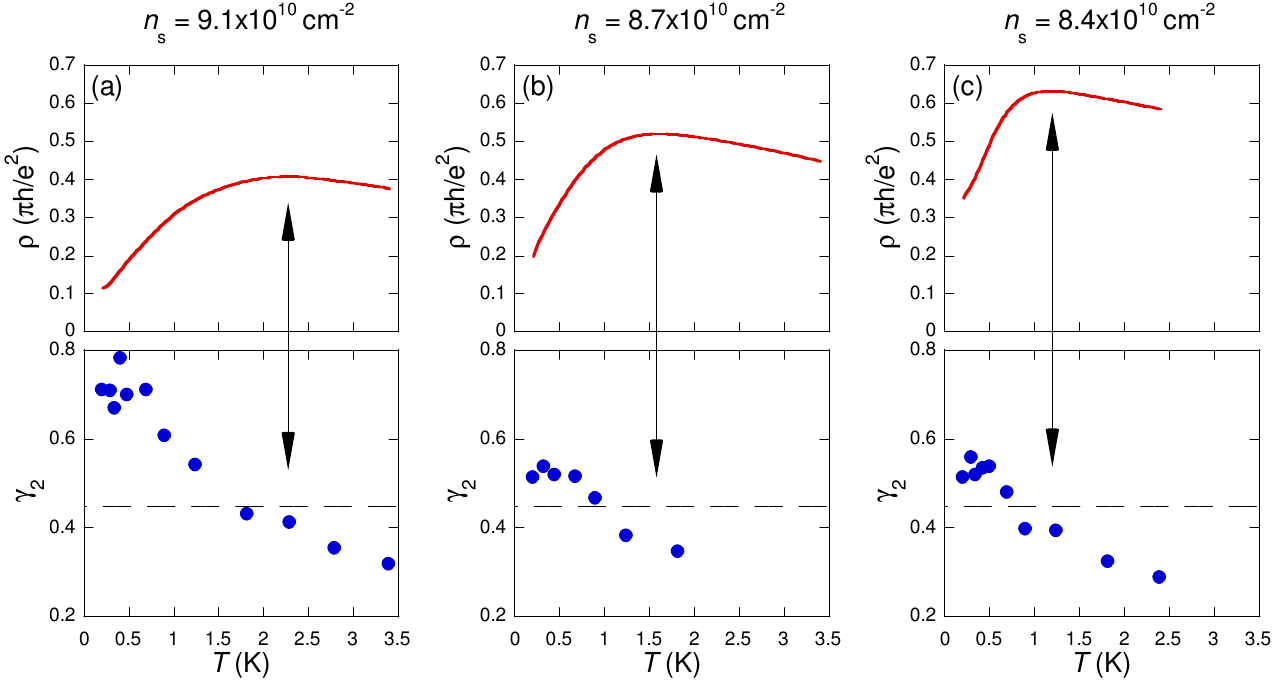}
        \caption{Upper panels: $\rho(T)$ traces (in units of $\pi h/e^2$)
        for three different electron densities, $n_s=9.87,9.58$ and $9.14\times 10^{10}$\,cm$^{-2}$.
        Lower panels: Extracted values of $\gamma_2(T)$ using Eq.\,\ref{eqn:linear}
        using $K=4$, for the same electron densities. (See Ref.\,\cite{punnoose07} for further details.)
        The dashed lines are positioned at the critical value $\gamma_2^*=0.45$.
        Note that the maximum in $\rho(T)$ occurs when $\gamma_2$ attains approximately
        the value $\gamma_2^*$.
        }
    \label{fig:rawdata}
\end{figure}

\begin{figure}[htb]
    \centering
        \includegraphics[width=0.4\textwidth]{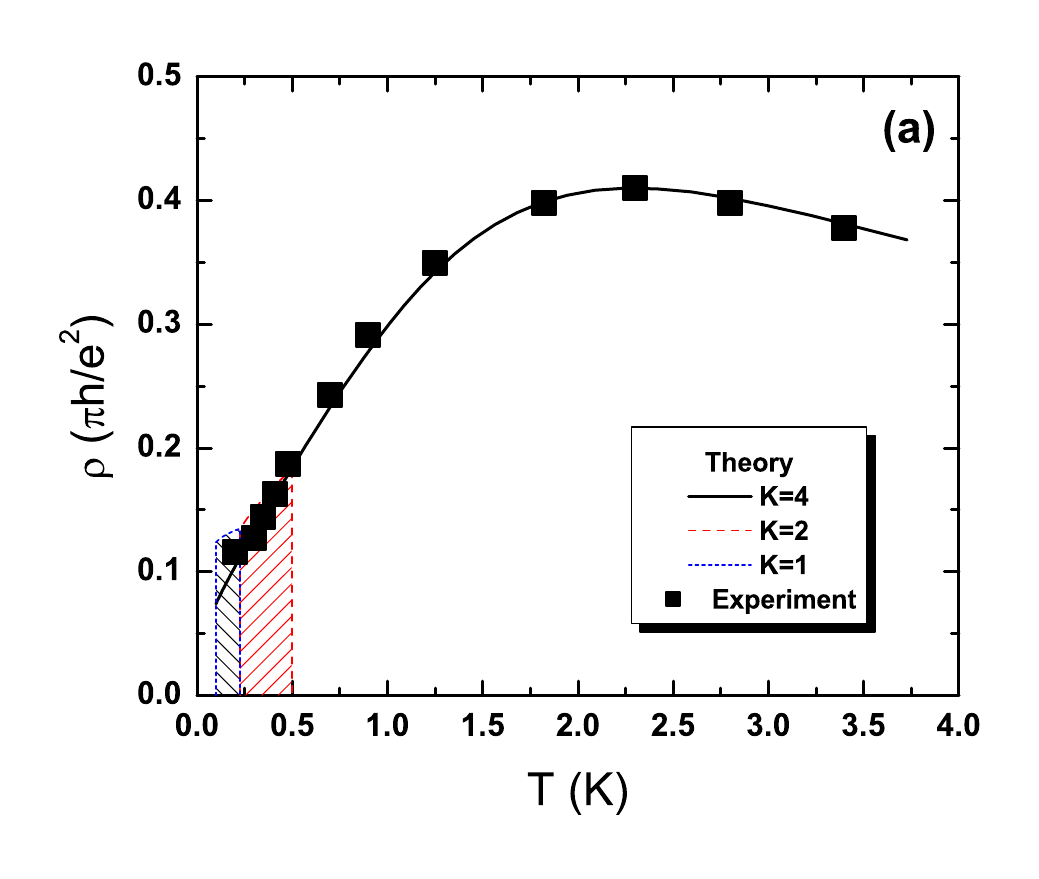}
        \includegraphics[width=0.4\textwidth]{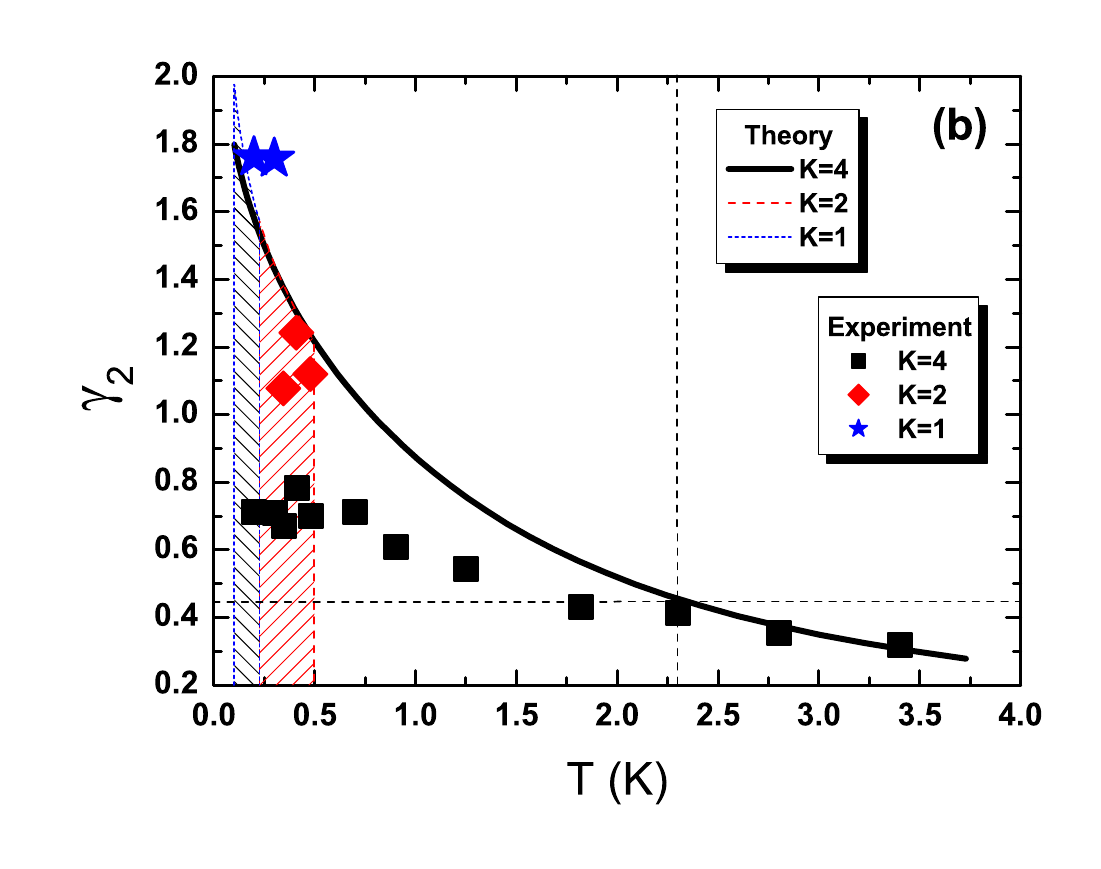}
        \caption{The result of the comparison between theory (lines)
        and experiment (symbols) for $\rho$ and $\gamma_2$ are presented
        in (a) and (b), respectively. The parameter $K=4$ corresponds
        to the case when the two valleys ($n_v=2$) are degenerate, i.e.,
        $T> T_v$, where $T_v\approx 0.5$\,K is the estimated valley splitting.
        $K=2$ corresponds to the temperature range $T\alt T_v$, and $K=1$
        corresponds to the region $T\alt T_{\perp}\approx 0.2$\,K where the
        intervalley scattering mixes the two valleys to give one valley.}
    \label{fig:fit}
\end{figure}

In the upper panels in Fig.\,\ref{fig:rawdata}, we plot $\rho(T)$ at zero
magnetic field for three different electron densities.  They show a
characteristic non-monotonic behavior as predicted in (\ref{eqn:RG}).
In the lower panels in Fig.\,\ref{fig:rawdata} we plot the extracted values of $\gamma_2$ using Eq.\,(\ref{eqn:linear}) with $K=4$, i.e., assuming that the valleys are degenerate and distinct. The dashed horizontal line marks the point $\gamma_2\approx 0.45$ approximately where $\rho(T)$ attains its maximum value in remarkable agreement with Eq.\,(\ref{eqn:RG}). (At these temperatures quantum coherence is relevant and its contribution to weak localization, $d\rho/d\xi=n_v\rho^2$, is to be added  to Eq.\,(\ref{eqn:RG}a).)
The results of the comparison between theory and experiment are
presented in Fig.\,\ref{fig:fit}. The solid squares ($\blacksquare$) are
the experimental data points for $n_s= 9.1\times 10^{10}$\,cm$^{-2}$,
reproduced here from Fig.\,\ref{fig:rawdata}(a). The solid lines  are the
predicted theoretical curves for $\rho(T)$ and $\gamma_2(T)$ with
the parameters $K=4$, $\rho_{max}=0.4$ and $T_{max}=2.3$\,K. (Here, $T_{max}$ is the temperature at which $\rho(T)$ attains its maximum value, $\rho(T_{max})=\rho_{max}$.) The
remarkable agreement between theory and experiment is especially
striking given that the theory has no adjustable parameters.

At temperatures below  0.5\,K, the experimentally extracted values of
$\gamma_2(T)$ in Fig.\,\ref{fig:fit}(b) seem to saturate with further
decrease in $T$. We believe that the saturation is an artifact of the analysis
related to our assumptions that both the valley splitting and the
intervalley scattering are negligible at the  lowest
temperatures. As noted  earlier, the large number of valley modes $K=n_v^2$ reduces to just $K=n_v$ for temperature $T_\perp\alt T\alt T_v$ and to just $K=1$ for $T\alt T_\perp$. In the following, we recalculate $\gamma_2(T)$ taking these considerations into account. 

The experimentally extracted values of $\gamma_2$, using $K=2$ and
$K=1$, are shown in Fig.~\ref{fig:fit}(b) as diamonds (red
{\color{red}$\blacklozenge$}) and stars (blue {\color{blue}$\bigstar$}), respectively. The
procedure used to extract these values are the same as that used for
$K=4$, namely, by fitting the $\sigma(b^2)$ traces in
Fig.~\ref{fig:rawdata}(b) to Eq.~(\ref{eqn:linear}) using the appropriate
$K$ values. We find very favorable agreement with theory (solid line)  if the
crossover scales are chosen such that  $T_v\approx 0.5$\,K and
$T_{\perp}\approx 0.2$\,K. (Note that for these temperatures the WL corrections have not been included in Eq.\,(\ref{eqn:RG}) for the reasons discussed earlier.) These values are in good agreement with earlier estimates of $T_v$\,\cite{ZNAfits_vitkalov} and $T_\perp$\,\cite{tau_perp_gershenson} obtained at higher densities employing different methods. 
We checked by direct calculation using Eq.\,(\ref{eqn:RG}) that the
theoretical values of $\rho$ and $\gamma_2$ are  not affected
significantly when crossing these scales, provided that the WL corrections are not included below $T\alt 500$K.  Deviations from the solution for $K=4$ taking $K=2$ and $K=1$ are shown in
Fig.~\ref{fig:fit} as long (red) and short (blue) dashed lines,
respectively. As can be seen, the deviations are insignificant (almost indiscernible) down
to $T=0.2$\,K. 

By comparing with experiments we have extended the test of the scaling equations (\ref{eqn:RG}) down to the lowest reliably measurable temperatures $T\approx 0.2$\,K. Concerning still lower temperatures, i.e., lower than $T= 0.2$\,K, the theory predicts (not shown here) that while $\rho(T)$ saturates and then
begins to drop again at ultra low temperatures ($T\alt 50$\,mK),
$\gamma_2(T)$ rises fast monotonically  for
$K=1$. Further  tests of these predictions  are in progress. 

To conclude, we have shown that if  valley splitting and intervalley scattering are incorporated into the RG theory, the latter quantitatively describes the metallic phase down to the lowest readily accessible temperatures. The extracted values of intervalley scattering time and valley splitting are in good agreement with those previously obtained at higher densities using different methods.  

\begin{thebibliography}{17}
\expandafter\ifx\csname natexlab\endcsname\relax\def\natexlab#1{#1}\fi
\expandafter\ifx\csname bibnamefont\endcsname\relax
  \def\bibnamefont#1{#1}\fi
\expandafter\ifx\csname bibfnamefont\endcsname\relax
  \def\bibfnamefont#1{#1}\fi
\expandafter\ifx\csname citenamefont\endcsname\relax
  \def\citenamefont#1{#1}\fi
\expandafter\ifx\csname url\endcsname\relax
  \def\url#1{\texttt{#1}}\fi
\expandafter\ifx\csname urlprefix\endcsname\relax\def\urlprefix{URL }\fi
\providecommand{\bibinfo}[2]{#2}
\providecommand{\eprint}[2][]{\url{#2}}

\bibitem[{\citenamefont{Punnoose and Finkel'stein}(2002)}]{punnoose02}
\bibinfo{author}{\bibfnamefont{A.}~\bibnamefont{Punnoose}} \bibnamefont{and}
  \bibinfo{author}{\bibfnamefont{A.~M.} \bibnamefont{Finkel'stein}},
  \bibinfo{journal}{Phys. Rev. Lett.} \textbf{\bibinfo{volume}{88}},
  \bibinfo{pages}{016802} (\bibinfo{year}{2002}).

\bibitem[{\citenamefont{Punnoose and Finkel'stein}(2005)}]{punnoose05}
\bibinfo{author}{\bibfnamefont{A.}~\bibnamefont{Punnoose}} \bibnamefont{and}
  \bibinfo{author}{\bibfnamefont{A.~M.} \bibnamefont{Finkel'stein}},
  \bibinfo{journal}{Science} \textbf{\bibinfo{volume}{310}},
  \bibinfo{pages}{289} (\bibinfo{year}{2005}).

\bibitem[{\citenamefont{Anissimova et~al.}(2007)\citenamefont{Anissimova,
  Kravchenko, Punnoose, Finkel'stein, and Klapwijk}}]{punnoose07}
\bibinfo{author}{\bibfnamefont{S.}~\bibnamefont{Anissimova}},
  \bibinfo{author}{\bibfnamefont{S.~V.} \bibnamefont{Kravchenko}},
  \bibinfo{author}{\bibfnamefont{A.}~\bibnamefont{Punnoose}},
  \bibinfo{author}{\bibfnamefont{A.~M.} \bibnamefont{Finkel'stein}},
  \bibnamefont{and} \bibinfo{author}{\bibfnamefont{T.~M.}
  \bibnamefont{Klapwijk}}, \bibinfo{journal}{Nature Physics}
  \textbf{\bibinfo{volume}{3}}, \bibinfo{pages}{707} (\bibinfo{year}{2007}).

\bibitem[{\citenamefont{Knyazev et~al.}(2006)\citenamefont{Knyazev,
  Omel'yanovskii, Pudalov, and Burmistrov}}]{pudalov_RG_JETP}
\bibinfo{author}{\bibfnamefont{D.~A.} \bibnamefont{Knyazev}},
  \bibinfo{author}{\bibfnamefont{O.~E.} \bibnamefont{Omel'yanovskii}},
  \bibinfo{author}{\bibfnamefont{V.~M.} \bibnamefont{Pudalov}},
  \bibnamefont{and} \bibinfo{author}{\bibfnamefont{I.~S.}
  \bibnamefont{Burmistrov}}, \bibinfo{journal}{JETP Lett.}
  \textbf{\bibinfo{volume}{84}}, \bibinfo{pages}{662} (\bibinfo{year}{2006}).

\bibitem[{\citenamefont{Knyazev et~al.}(2008)\citenamefont{Knyazev,
  Omel'yanovskii, Pudalov, and Burmistrov}}]{pudalov_RG_PRL}
\bibinfo{author}{\bibfnamefont{D.~A.} \bibnamefont{Knyazev}},
  \bibinfo{author}{\bibfnamefont{O.~E.} \bibnamefont{Omel'yanovskii}},
  \bibinfo{author}{\bibfnamefont{V.~M.} \bibnamefont{Pudalov}},
  \bibnamefont{and} \bibinfo{author}{\bibfnamefont{I.~S.}
  \bibnamefont{Burmistrov}}, \bibinfo{journal}{Phys. Rev. Lett.}
  \textbf{\bibinfo{volume}{100}}, \bibinfo{pages}{046405}
  (\bibinfo{year}{2008}).

\bibitem[{\citenamefont{Finkel'stein}(1990)}]{yellowbook}
\bibinfo{author}{\bibfnamefont{A.~M.} \bibnamefont{Finkel'stein}},
  \bibinfo{journal}{Sov. Sci. Rev. A, Phys. Rev.}
  \textbf{\bibinfo{volume}{14}}, \bibinfo{pages}{1} (\bibinfo{year}{1990}).

\bibitem[{\citenamefont{Castellani et~al.}(1984)\citenamefont{Castellani,
  Castro, Lee, and Ma}}]{pedagogical}
\bibinfo{author}{\bibfnamefont{C.}~\bibnamefont{Castellani}},
  \bibinfo{author}{\bibfnamefont{C.~D.} \bibnamefont{Castro}},
  \bibinfo{author}{\bibfnamefont{P.~A.} \bibnamefont{Lee}}, \bibnamefont{and}
  \bibinfo{author}{\bibfnamefont{M.}~\bibnamefont{Ma}}, \bibinfo{journal}{Phys.
  Rev. B} \textbf{\bibinfo{volume}{30}}, \bibinfo{pages}{527}
  (\bibinfo{year}{1984}).

\bibitem[{\citenamefont{Punnoose}()}]{punnoose09a}
\bibinfo{author}{\bibfnamefont{A.}~\bibnamefont{Punnoose}},
  \bibinfo{howpublished}{arXiv:0910.0037}.

\bibitem[{\citenamefont{Ando et~al.}(1982)\citenamefont{Ando, Fowler, and
  Stern}}]{ando}
\bibinfo{author}{\bibfnamefont{T.}~\bibnamefont{Ando}},
  \bibinfo{author}{\bibfnamefont{A.~B.} \bibnamefont{Fowler}},
  \bibnamefont{and} \bibinfo{author}{\bibfnamefont{F.}~\bibnamefont{Stern}},
  \bibinfo{journal}{Rev. Mod. Phys.} \textbf{\bibinfo{volume}{54}},
  \bibinfo{pages}{437} (\bibinfo{year}{1982}).

\bibitem[{\citenamefont{Fukuyama}(1980)}]{fukuyama1}
\bibinfo{author}{\bibfnamefont{H.}~\bibnamefont{Fukuyama}},
  \bibinfo{journal}{J. Phys. Soc. Jpn.} \textbf{\bibinfo{volume}{49}},
  \bibinfo{pages}{649} (\bibinfo{year}{1980}).

\bibitem[{\citenamefont{Burmistrov and
  Chtchelkatchev}(2008)}]{burmistrov_spinvalley}
\bibinfo{author}{\bibfnamefont{I.~S.} \bibnamefont{Burmistrov}}
  \bibnamefont{and} \bibinfo{author}{\bibfnamefont{N.~M.}
  \bibnamefont{Chtchelkatchev}}, \bibinfo{journal}{Phys. Rev. B}
  \textbf{\bibinfo{volume}{77}}, \bibinfo{pages}{195319}
  (\bibinfo{year}{2008}).

\bibitem[{\citenamefont{Altshuler and Aronov}(1985)}]{aabook}
\bibinfo{author}{\bibfnamefont{B.~L.} \bibnamefont{Altshuler}}
  \bibnamefont{and} \bibinfo{author}{\bibfnamefont{A.~G.}
  \bibnamefont{Aronov}}, \emph{\bibinfo{title}{{Modern Problems in Condensed
  Matter Physics}}} (\bibinfo{publisher}{Elsevier, North Holland},
  \bibinfo{year}{1985}), chap. \bibinfo{chapter}{Electron-Electron Interactions
  in Disordered Systems}, p.~\bibinfo{pages}{1}.

\bibitem[{\citenamefont{Rahimi et~al.}(2003)\citenamefont{Rahimi, Anissimova,
  Sakr, and Kravchenko}}]{dephasing_kravchenko}
\bibinfo{author}{\bibfnamefont{M.}~\bibnamefont{Rahimi}},
  \bibinfo{author}{\bibfnamefont{S.}~\bibnamefont{Anissimova}},
  \bibinfo{author}{\bibfnamefont{M.~R.} \bibnamefont{Sakr}}, \bibnamefont{and}
  \bibinfo{author}{\bibfnamefont{S.~V.} \bibnamefont{Kravchenko}},
  \bibinfo{journal}{Phys. Rev. Lett.} \textbf{\bibinfo{volume}{91}},
  \bibinfo{pages}{116402} (\bibinfo{year}{2003}).

\bibitem[{\citenamefont{Lee and Ramakrishnan}(1982)}]{tvr}
\bibinfo{author}{\bibfnamefont{P.~A.} \bibnamefont{Lee}} \bibnamefont{and}
  \bibinfo{author}{\bibfnamefont{T.~V.} \bibnamefont{Ramakrishnan}},
  \bibinfo{journal}{Phys. Rev. B} \textbf{\bibinfo{volume}{26}},
  \bibinfo{pages}{4009} (\bibinfo{year}{1982}).

\bibitem[{\citenamefont{Castellani et~al.}(1998)\citenamefont{Castellani,
  Castro, and Lee}}]{castellani98}
\bibinfo{author}{\bibfnamefont{C.}~\bibnamefont{Castellani}},
  \bibinfo{author}{\bibfnamefont{C.~D.} \bibnamefont{Castro}},
  \bibnamefont{and} \bibinfo{author}{\bibfnamefont{P.~A.} \bibnamefont{Lee}},
  \bibinfo{journal}{Phys. Rev. B} \textbf{\bibinfo{volume}{57}},
  \bibinfo{pages}{R9381} (\bibinfo{year}{1998}).

\bibitem[{\citenamefont{Vitkalov et~al.}(2003)\citenamefont{Vitkalov, James,
  Narozhny, Sarachik, and Klapwijk}}]{ZNAfits_vitkalov}
\bibinfo{author}{\bibfnamefont{S.~A.} \bibnamefont{Vitkalov}},
  \bibinfo{author}{\bibfnamefont{K.}~\bibnamefont{James}},
  \bibinfo{author}{\bibfnamefont{B.~N.} \bibnamefont{Narozhny}},
  \bibinfo{author}{\bibfnamefont{M.~P.} \bibnamefont{Sarachik}},
  \bibnamefont{and} \bibinfo{author}{\bibfnamefont{T.~M.}
  \bibnamefont{Klapwijk}}, \bibinfo{journal}{Phys. Rev. B}
  \textbf{\bibinfo{volume}{67}}, \bibinfo{pages}{113310}
  (\bibinfo{year}{2003}).

\bibitem[{\citenamefont{Kuntsevich et~al.}(2007)\citenamefont{Kuntsevich,
  Klimov, Tarasenko, Averkiev, Pudalov, Kojima, and
  Gershenson}}]{tau_perp_gershenson}
\bibinfo{author}{\bibfnamefont{A.~Y.} \bibnamefont{Kuntsevich}},
  \bibinfo{author}{\bibfnamefont{N.~N.} \bibnamefont{Klimov}},
  \bibinfo{author}{\bibfnamefont{S.~A.} \bibnamefont{Tarasenko}},
  \bibinfo{author}{\bibfnamefont{N.~S.} \bibnamefont{Averkiev}},
  \bibinfo{author}{\bibfnamefont{V.~M.} \bibnamefont{Pudalov}},
  \bibinfo{author}{\bibfnamefont{H.}~\bibnamefont{Kojima}}, \bibnamefont{and}
  \bibinfo{author}{\bibfnamefont{M.~E.} \bibnamefont{Gershenson}},
  \bibinfo{journal}{Phys. Rev. B} \textbf{\bibinfo{volume}{75}},
  \bibinfo{pages}{195330} (\bibinfo{year}{2007}).

\end{thebibliography}

\end{document}